\documentclass[11pt]{article}
\usepackage{pdfpages}
\usepackage{graphics}
\usepackage{wrapfig,rotating}
\usepackage[dvips]{epsfig}
\def\be{\begin{equation}}
\def\ee{\end{equation}}
\def\bqa{\begin{eqnarray}}
\def\eqa{\end{eqnarray}}
\def\lsim{\raise0.3ex\hbox{$<$\kern-0.75em\raise-1.1ex\hbox{$\sim$}}}
\def\gsim{\raise0.3ex\hbox{$>$\kern-0.75em\raise-1.1ex\hbox{$\sim$}}}
\usepackage{amssymb}

\begin{document}

\begin{center}
{\Large\bf The Color Dipole Picture\footnote{Presented at Diffraction 2012,
Lanzarote, Canary Islands (Spain), September 10-15, 2012 (Proceedings
to appear)}}
\vspace*{0.5cm}

Dieter Schildknecht
\vspace*{0.5cm}

{\footnotesize\sl Fakult\"at f\"ur Physik, Universit\"at Bielefeld,\\[1.2mm]
  Universit\"atsstra\ss e 25, 33615 Bielefeld, Germany\\[1.2mm]
and\\[1.2mm]
Max-Planck-Institute for Physics, F\"ohringer Ring 6, \\[1.2mm]
80805 Munich, Germany}
\end{center}
\vspace*{0.5cm}

{\footnotesize{\bf Abstract.} 
We give a brief exposition of the color dipole picture of deep
inelastic scattering.}\vspace*{0.2cm}



\begin{center}{\bf INTRODUCTION}\end{center}

\noindent
In Fig. 1 \cite{PRD}, I show\footnote{For a more elaborate presentation of the color
dipole picture and a more complete list of references compare also
the recent review papers in refs. \cite{Ringberg} and \cite{Erice}.}
 the experimental data for the proton electromagnetic structure function
$F_2 (x, Q^2)$ as a function of $1/W^2$, and, for comparison, as a function
of the Bjorken variable $x = Q^2/(W^2 + Q^2-M^2) \simeq Q^2/W^2~
({\rm for}~x \ll 0.1)$. For photon virtualities of $10~ {\rm GeV}^2 \le Q^2 \le
100~{\rm GeV}^2$, and for the photon-proton center of mass energy, $W$, sufficiently
large, we observe a simple scaling behavior $F_2 (x,Q^2) = F_2 (W^2)$. 
\begin{figure}[h]
\includegraphics[scale=.35]{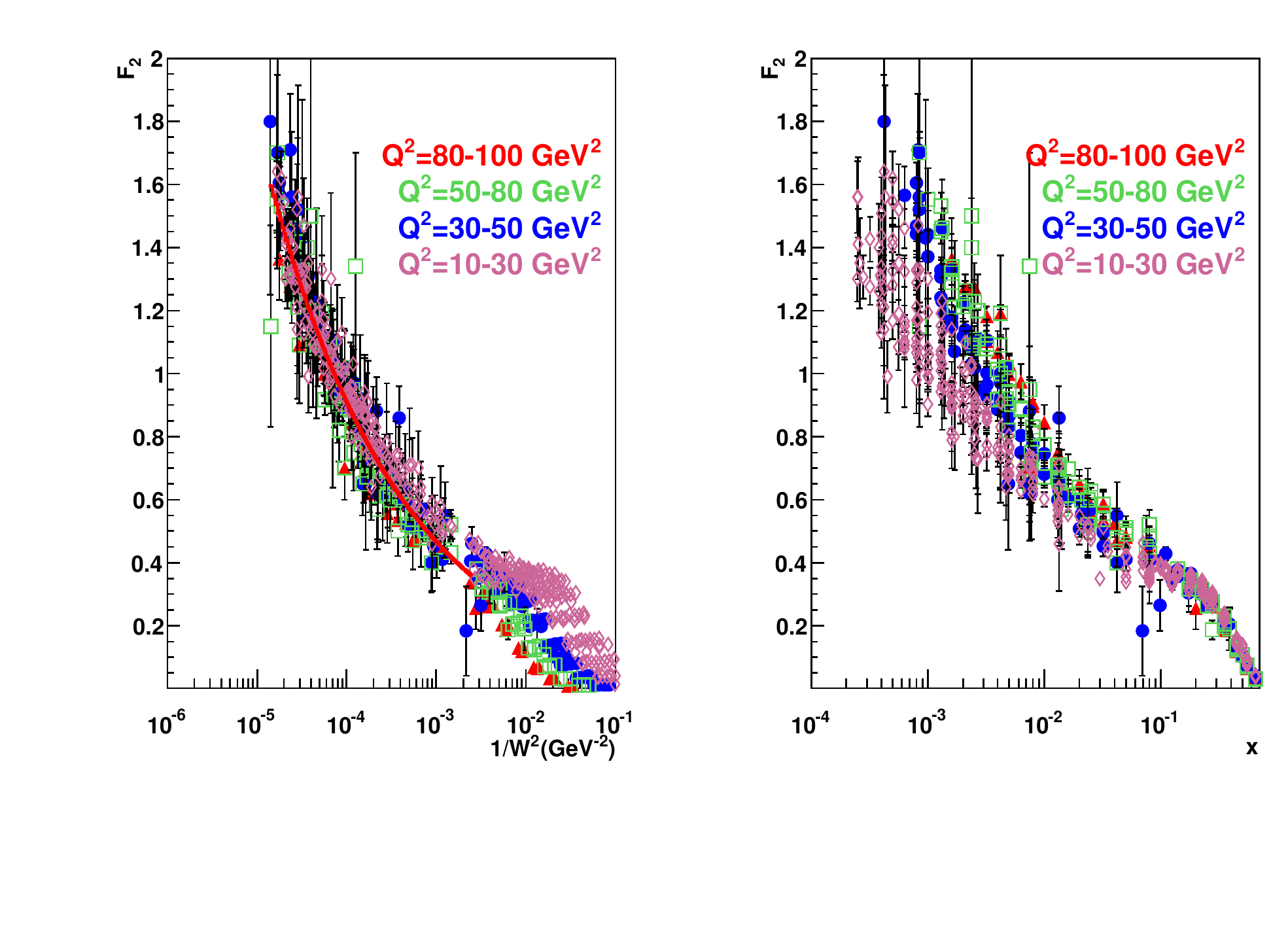}
\hspace*{1cm}
\includegraphics[scale=.25]{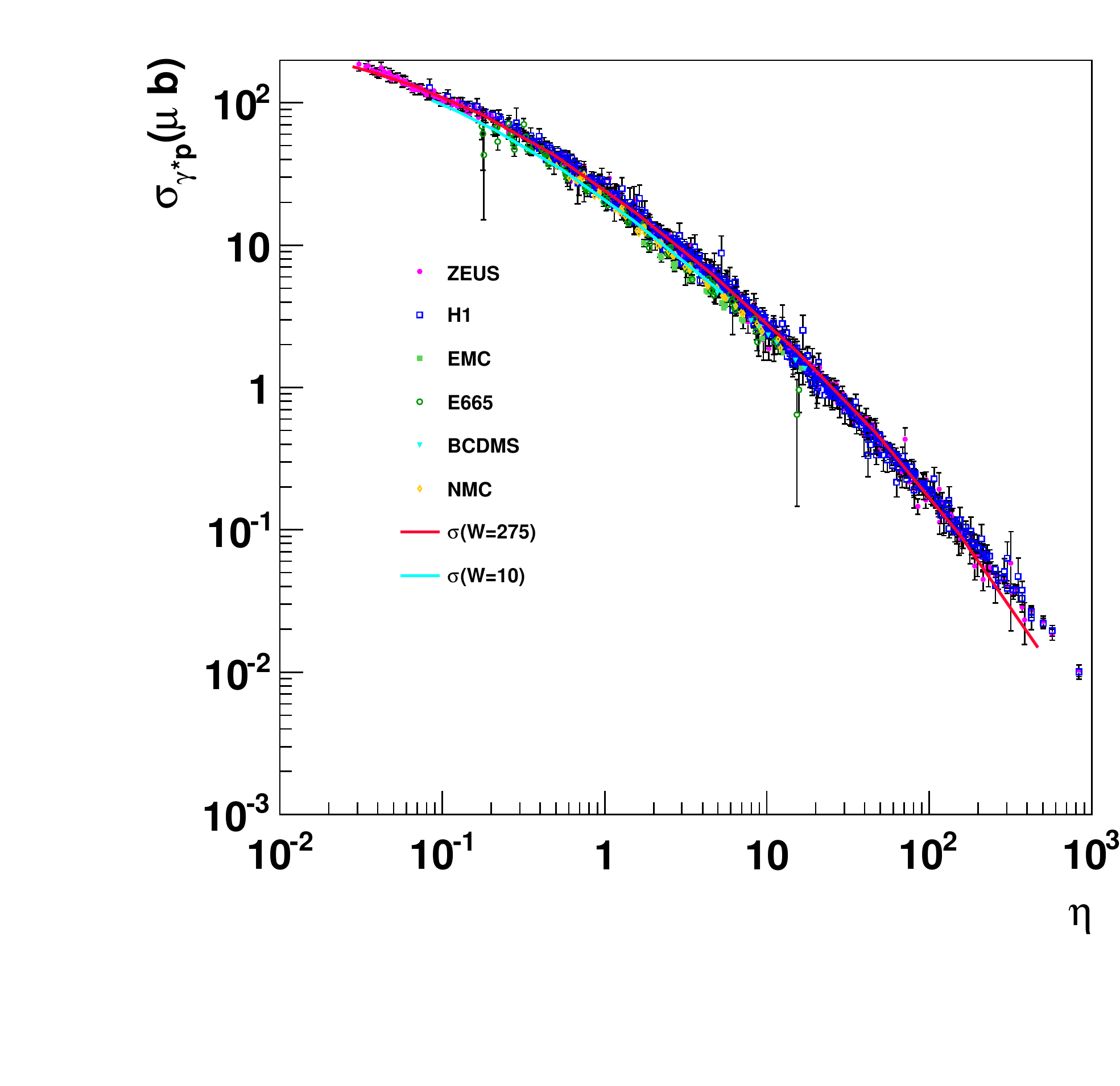}
\vspace*{-1cm}
\caption{\hspace*{-0.5cm} {\footnotesize The structure function $F_2 (x, Q^2)$.}
\hspace*{0.5cm} 
Figure 2:
{\footnotesize Scaling \cite{DIFF2000,SCHI} of 
\hspace*{9cm}$\sigma_{\gamma^*p} (W^2,Q^2)
= \sigma_{\gamma^*p}(\eta (W^2,Q^2)).$}}
\end{figure}
The
theoretical curve \cite{PRD} in Fig. 1 is based on $F_2(W^2) = 
f_2 (W^2/1~{\rm GeV}^2)^{C_2}$, \break
where $f_2 = 0.063$ and $C_2 = 0.29$. In terms
of the photoabsorption cross section, $\sigma_{\gamma^*p} (W^2,Q^2) \simeq
(4 \pi^2 \alpha/Q^2) F_2 (x,Q^2)$, upon introducing \cite{DIFF2000}  
the low-x scaling 
variable 
\be
\eta (W^2,Q^2) = \frac{Q^2 + m^2_0}{\Lambda^2_{sat} (W^2)},
\label{1}
\ee
where $m^2_0 \cong 0.15~{\rm GeV}^2$ and $\Lambda^2_{sat} (W^2) \sim
(W^2/1~{\rm GeV}^2)^{C_2}$, one finds scaling \cite{DIFF2000, SCHI} in an
extended region that includes the transition to $Q^2 \to 0$. 
Compare
Fig. 2. From Fig. 2, we read off the functional dependence of,
\be
\sigma_{\gamma^*p} (W^2,Q^2) = \sigma_{\gamma^*p} (\eta(W^2,Q^2)) \sim
\sigma^{(\infty)}
\left\{ \matrix{
\frac{1}{\eta (W^2, Q^2)},~~~~~~{\rm for}~~(\eta (W^2, Q^2) \gg 1), \cr
\ln \frac{1}{\eta (W^2,Q^2)},
~~~{\rm for}~~ (\eta (W^2,Q^2) \ll 1), } \right.
\label{2}
\ee
where the cross section $\sigma^{(\infty)}$ is at most very weakly
dependent on the energy $W$. 

\setcounter{figure}{2}
\begin{figure}[t]
\vspace*{-2cm}
\hspace*{1cm}
\includegraphics[scale=.25]{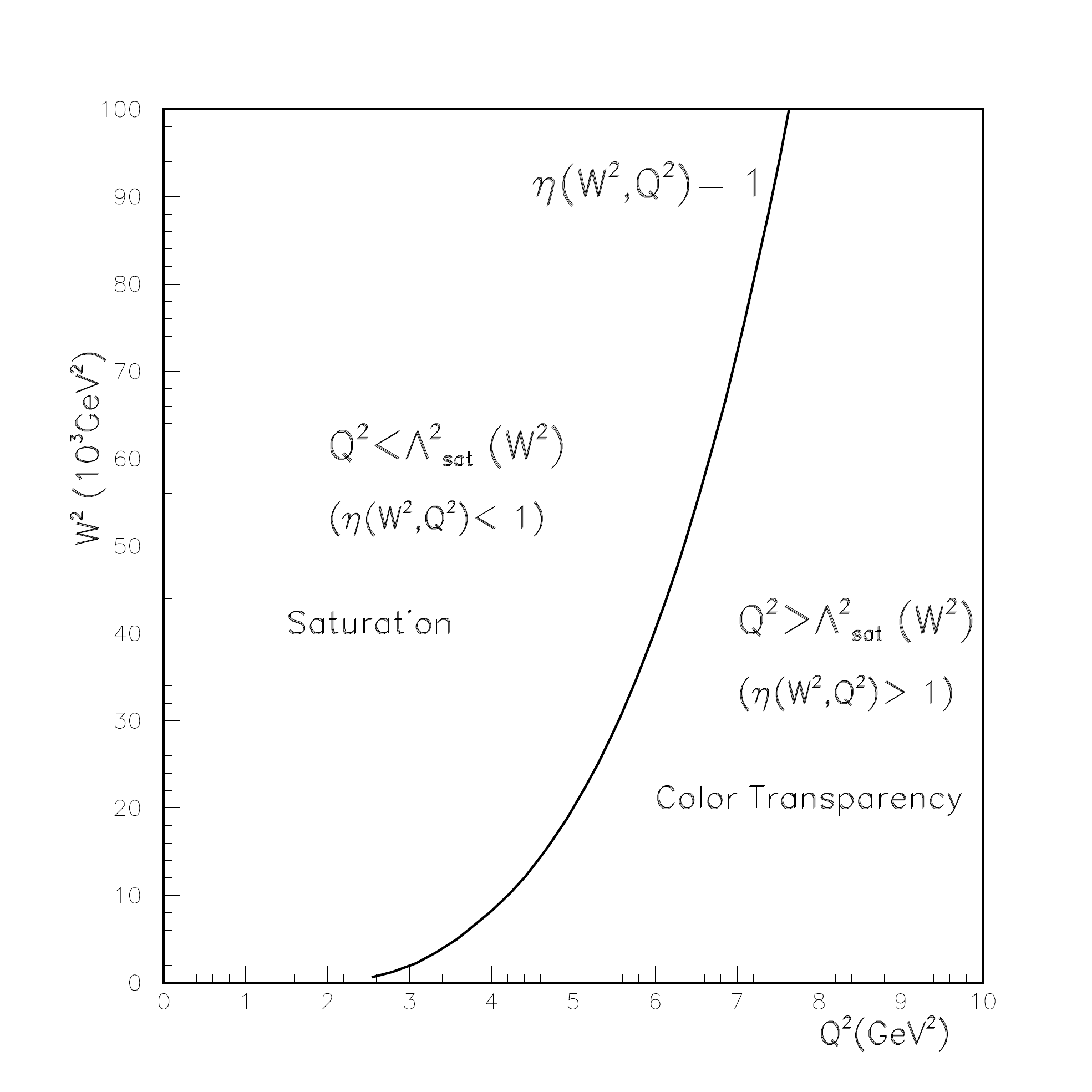}\hspace*{2cm}
\includegraphics[scale=.25]{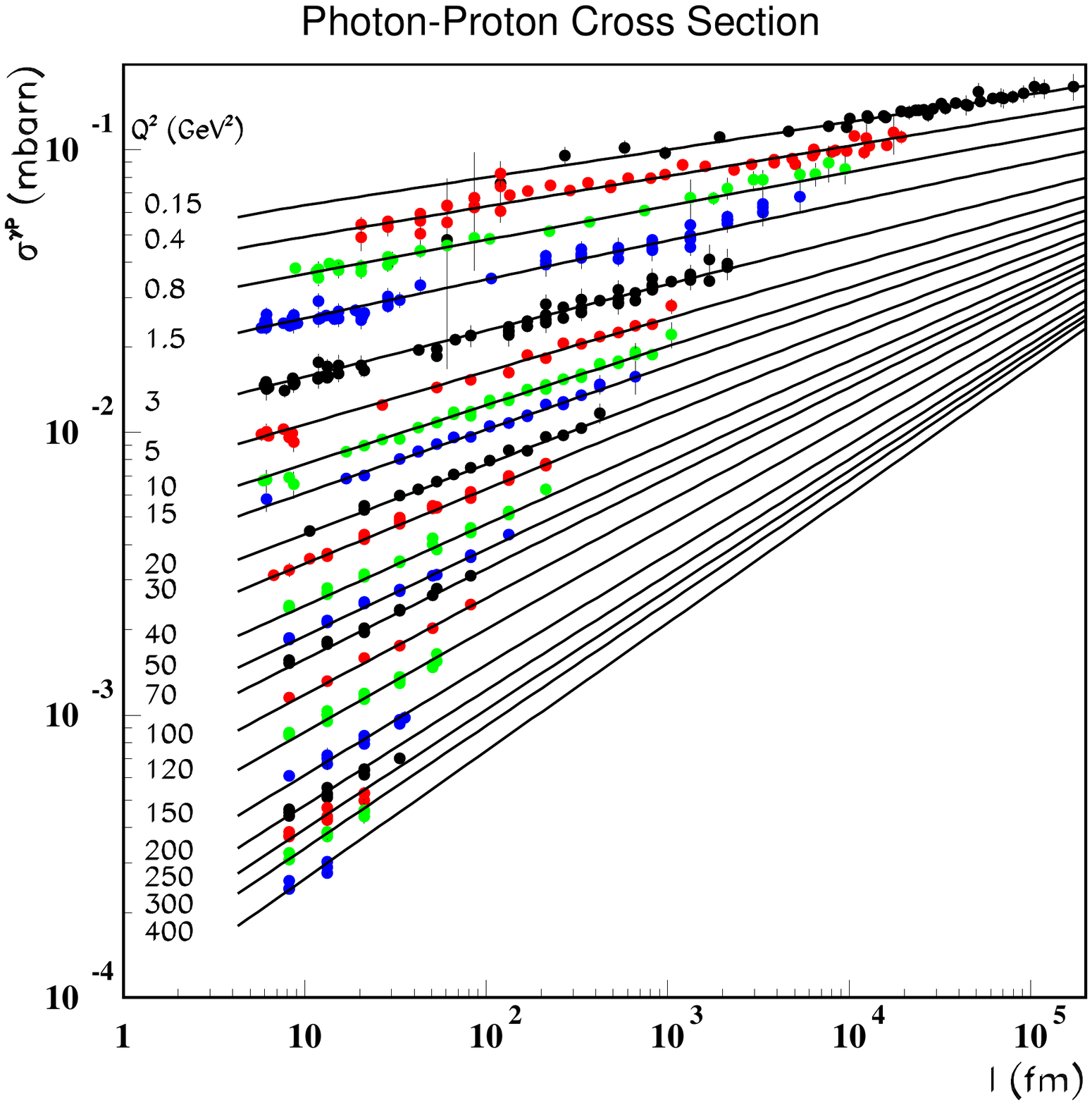}
\caption{{\footnotesize The $(Q^2,W^2)$ plane.}\hspace*{2cm} 
{Figure 4:} {\footnotesize The Caldwell fit\cite{CAL}.}}
\end{figure}

The $(Q^2,W^2)$ plane corresponding to the 
behavior in Fig. 2 is simple, compare Fig. 3. The logarithmic dependence
in (\ref{2}) implies the transition from the $1/\eta (W^2,Q^2)$ 
dependence of ``color transparency'' to the $Q^2$-independent 
``saturation'' limit that coincides \cite{DIFF2000, SCHI} with $Q^2 = 0$
photoproduction
\be
\lim_{W^2 \to \infty \atop {Q^2~{\rm fixed}}} 
\frac{\sigma_{\gamma^*p} (\eta (W^2, Q^2))}{\sigma_{\gamma^*p} (\eta
(W^2, Q^2 = 0))} 
= \lim_{W^2 \to \infty \atop {Q^2~{\rm fixed}}} 
\frac{\ln \left( \frac{\Lambda^2_{sat}(W^2)}{m^2_0} 
\frac{m^2_0}{(Q^2 + m^2_0)} \right)}{\ln 
\frac{\Lambda^2_{sat} (W^2)}{m^2_0}} 
= 1 + \lim_{W^2 \to \infty \atop {Q^2~{\rm fixed}}} 
\frac{\ln \frac{m^2_0}{Q^2 + m^2_0}}{\ln 
\frac{\Lambda^2_{sat} (W^2)}{m^2_0}} = 1.
\label{3}
\ee
The approach to a $Q^2$-independent limit for $W^2 \to \infty$ at
$Q^2$ fixed was recently also observed by Caldwell \cite{CAL}, compare Fig. 4, 
showing the empirical fit of $\sigma_{\gamma^*p} (W^2,Q^2) = \sigma_0 (Q^2)
(W^2/2M_pQ^2)^{\lambda_{eff} (Q^2)}$.

In what follows, I shall point out that not only the existence of scaling
in $\eta (W^2,Q^2)$, but the specific dependence on $\eta (W^2,Q^2)$ in
(\ref{2}) as well, both are unique and general model-independent
consequences from the color dipole picture (CDP).

\begin{center}{\bf THE COLOR DIPOLE PICTURE: MODEL-INDEPENDENT CONCLUSIONS}
\end{center}

\noindent
At low values of $x \ll 0.1$, deep inelastic scattering proceeds via
interaction of long-lived quark-antiquark, $q \bar q$, fluctuations of
the (virtual) photon that interact with the gluon field in the nucleon.
The total photoabsorption cross section is given by 
\cite{Nikolaev, Cvetic, DIFF2000, SCHI, PRD}
\bqa
\sigma_{\gamma^*_{L,T}p} (W^2, Q^2)  &=& \int dz \int d^2 \vec r_\bot
\vert \psi_{L,T} (\vec r_\bot, z (1 - z), Q^2) \vert^2 
\sigma_{(q\bar q)p}
(\vec r_\bot, z (1 - z), W^2) = \nonumber \\ 
&=&\frac{\alpha}{\pi} \sum_q Q^2_q Q^2 
\int dr^{\prime 2}_\bot K^2_{0,1} (r^\prime_\bot Q) 
\sigma_{(q \bar q)^{J=1}_{L,T} p} (r^{\prime 2}_\bot, W^2).
\label{4}
\eqa
In the second step in (\ref{4}), the explicit form of the square of the
``photon wave function'' $\vert \psi_{L,T} (\vec r_\bot, z (1-z),Q^2 \vert^2$
for massless quarks in terms of the modified Bessel functions $K^2_{0,1}
(r^\prime_\bot Q^2)$ was inserted, and a $J=1$ partial wave projection of the
$(q \bar q)p$ dipole cross section was performed. The transverse size
$\vec r^{~\prime}_\bot$ of the $(q \bar q)^{J=1}$ dipole states
is related to the transverse size
$\vec r_\bot$ via $\vec r^{~\prime}_\bot = \vec r_\bot \sqrt{z(1-z)}$ with
$0 \le z \le 1$ denoting the quark longitudinal momentum fraction of the
photon transition to $q \bar q$ pairs. The sum over the active quark charges
squared in (\ref{4}) is denoted by $\sum_q Q^2_q$.
\setcounter{figure}{4}
\begin{figure}[h]
\begin{center}
\includegraphics[scale=.6]{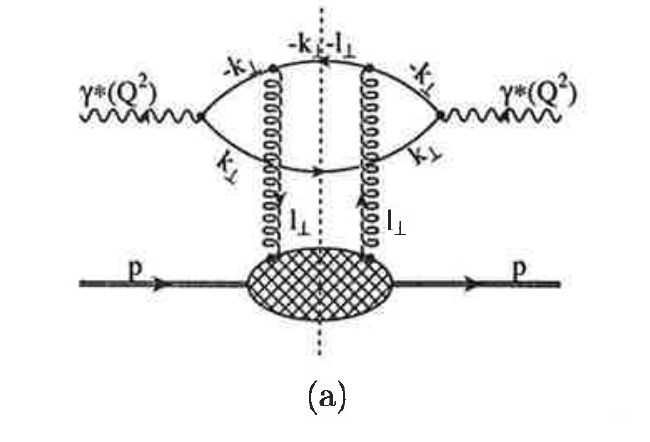}
\includegraphics[scale=.6]{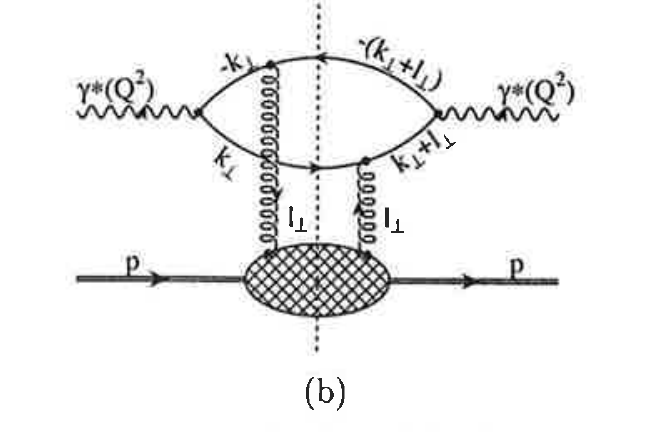}
\end{center}
\vspace*{-1cm}
\caption{{\footnotesize Two of the four diagrams for the $q \bar q$ 
dipole interaction.}}
\end{figure}

Two of the four contributing diagrams for the $q \bar q$ interaction with
the gluons in the nucleon are shown in Fig. 5. The color-gauge invariant
two-gluon interaction, as a consequence of the two diagrams in Fig. 5, 
implies a representation of the $(q \bar q)^{J=1}_{L,T} p$ cross section of
the form (e.g. ref. \cite{PRD})
\be
\sigma_{(q \bar q)^{J=1}_{L,T}p} (r^\prime_\bot, W^2)  = 
\pi \int d \vec l^{~\prime 2}_\bot \bar \sigma_{(q \bar q)^{J=1}_{L,T} p}
(\vec l^{~\prime 2}_\bot , W^2) 
( 1 - \frac{\int d \vec l^{~\prime 2}_\bot 
\bar \sigma_{(q \bar q)^{J=1}_{L,T} p} (\vec l^{~\prime 2}_\bot, W^2) J_0
(l^\prime_\bot r^\prime_\bot)}{\int d \vec l^{~\prime 2}_\bot
\bar \sigma_{(q \bar q)^{J=1}_{L,T} p} (\vec l^{~\prime 2}_\bot, W^2)}
), 
\label{5}
\ee
where $J_0(l^\prime_\bot r^\prime_\bot)$ denotes the Bessel function with
index $0$. At any fixed size $r^\prime_\bot$, we\break assume the dominant
contribution to the integrals in (\ref{5}) to be due to a restricted\break
range
of $r^\prime_\bot l^\prime_\bot \le r^\prime_\bot l^\prime_{\bot Max} (W^2)$,
with $\vec l^{~\prime}_{\bot Max}$ increasing with $W^2$. 
Inspection of the\break
integral in (\ref{5}) for fixed $r^\prime_\bot$ as a function of $W^2$,
reveals two different limits for the\break
dipole cross section on the left-hand 
side. There is either ``color transparency'', \break
$\sigma_{(q \bar q)^{J=1}_{L,T}}
(r^\prime_\bot, W^2) \sim r^{\prime 2}_\bot$ or else, ``saturation''
corresponding to the $r^\prime_\bot$-independent limit of 
$\sigma^{(\infty)}_{L,T} (W^2) = \pi \int d \vec l^{~\prime 2}_\bot
\bar \sigma_{(q \bar q)^{J=1}_{L,T} p} (\vec l^{~\prime 2}_\bot, W^2)$.
As a consequence of the strong fall-off of the modified Bessel functions in
(\ref{4}), the color transparency and saturation limits of the dipole cross
section translate into two different limits of the photoabsorption cross
section that are given by
\be
\sigma_{\gamma^*p} (W^2, Q^2)  =  \sigma_{\gamma^*p} (\eta
(W^2, Q^2)) = \frac{\alpha}{\pi} \sum_q Q^2_q  \cdot
\left\{ \matrix{
\frac{1}{6} ( 1 +2 \rho ) \sigma^{(\infty)}  \frac{1}{\eta(W^2, Q^2)},
 \cr
\sigma^{(\infty)}
\ln \frac{1}{\eta (W^2, Q^2)}, } \right.
\label{6}
\ee
where the upper and the lower line on the right hand side in (\ref{6}) refer
to $\eta (W^2,Q^2) \gg 1$ and $\eta (W^2,Q^2) \ll 1$, respectively.
The result (\ref{6}) from the CDP coincides  with the experimental one
(\ref{2}). In (\ref{6}), the ``saturation scale'', $\Lambda^2_{sat} (W^2)$
from (\ref{1}), is an increasing function of $W^2$ that is given by
$\Lambda^2_{sat} (W^2) = (\pi / \sigma^{(\infty)}_L) 
\int d \vec l^{~\prime 2}_{\bot} \vec l^{~\prime 2}_\bot 
\bar \sigma_{(q \bar q)^{J=1}_L p}(\vec l^{~\prime 2}_\bot, W^2)$, 
and $\rho$ is related to
the longitudinal-to-transverse ratio $R$ given by $R = 1/2 \rho$ at
sufficiently large $Q^2$, with $\rho = 4/3$ \cite{A, PRD}
due to the different size of
longitudinally-versus-transversely polarized $q \bar q$ states.

\begin{center}{\bf CONNECTION WITH THE PQCD-IMPROVED PARTON MODEL}
\end{center}

\vspace*{0.35cm}
\begin{wrapfigure}[19]{l}{0.48\textwidth}
\vspace*{-2.7cm}
\includegraphics[scale=.35]{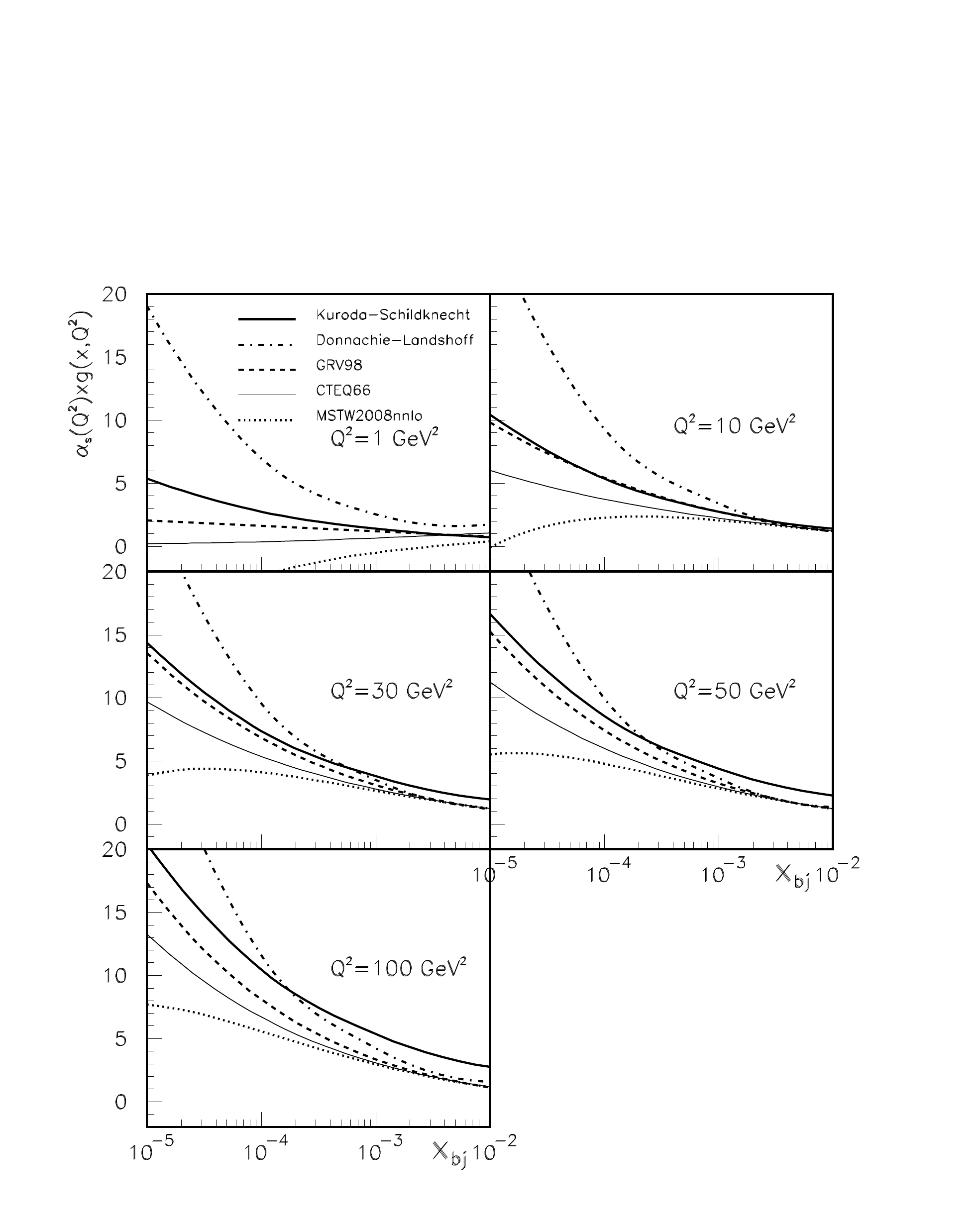}
\vspace*{-0.5cm}
\break{Figure 6:} \hspace*{-0.549736cm}
{\footnotesize
The gluon distribution function (\ref{8}) 
compared with
the results from the hard\hfill\break Pomeron part of a Regge fit 
\cite{Donnachie} and from\hfill\break several global pQCD fits \cite{DDB} to 
$F_2 (x,Q^2)$.}
\end{wrapfigure}
\noindent
Requiring consistency of the CDP with the pQCD-improved parton model, in
particular, requiring consistency with the evolution equation for the
structure function $F_2 (x,Q^2) \sim (W^2)^{C_2}$, as shown in Fig. 1,
one finds a simple explicit relation for the exponent $C_2$ that is given
by \cite{N9, PRD}
\be
C_2 = \frac{1}{2\rho + 1} \left( \frac{\xi_2}{\xi_L} \right)^{C_2} = 
0.29.
\label{7}
\ee 
The value of $C_2 = 0.29$ is consistent with the experimental results in
Figs. 1 and 2. In (\ref{7}) the previously mentioned predicted value of
$\rho = 4/3$ was inserted. The sensitivity on the rescaling factors in
the relevant interval of $1 \le \xi_2/\xi_L \le 1.5$ is fairly weak,
implying $0.27 \le C_2 \le 0.31$. 
The effective gluon distribution in
Fig. 6 \cite{PRD} based on
\be
\alpha_s (Q^2) G (x, Q^2)  
=\frac{3\pi}{\sum_q Q^2_q (2\rho + 1)} 
\frac{f_2}{\xi_L^{C_2 = 0.29}} \left(
  \frac{W^2}{1~ {\rm GeV}^2} \right)^{C_2 = 0.29}. 
\label{8}
\ee
is consistent with the widely varying results from the literature.
\newpage

\begin{center}{\bf THE COLOR DIPOLE PICTURE: MODEL-DEPENDENT PARAMETERIZATION}
\end{center}

\noindent
Any specific ansatz for the dipole cross section has to interpolate the
general model-independent functional dependences in (\ref{6}). For details
we have to refer to refs. \cite{DIFF2000, SCHI, PRD} and restrict ourselves
to showing the comparison \cite{PRD} with experiment in Figs. 7 and 8.
\setcounter{figure}{6}
\begin{figure}[h]
\begin{center}
\includegraphics[height=.3\textheight]{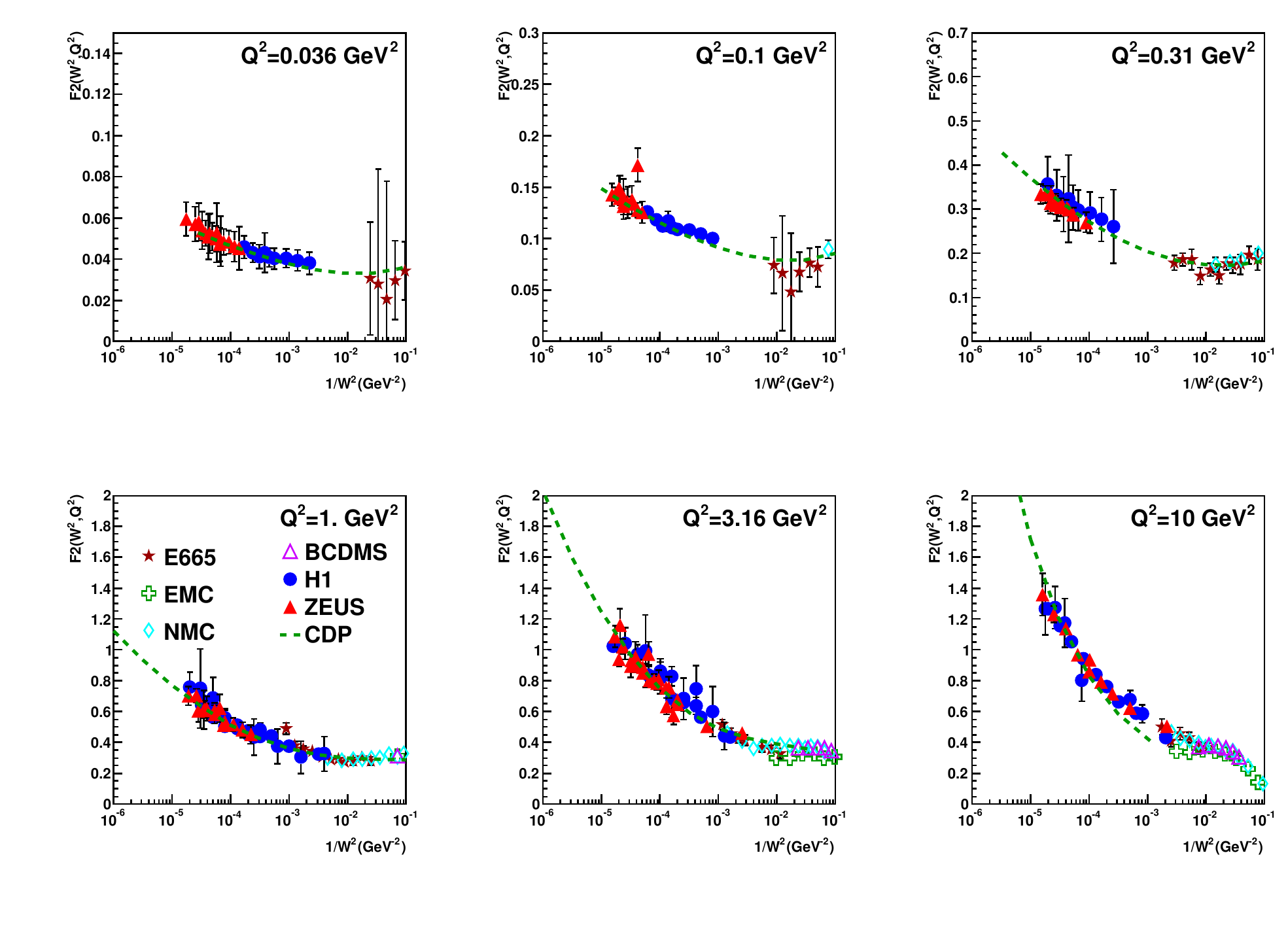}
\vspace*{-0.5cm}
\caption{{\footnotesize The predictions from the CDP \cite{PRD}
for the structure function
$F_2 (W^2,Q^2)$ compared with the experimental data for $0.036 \le Q^2 \le
10~{\rm GeV}^2$.}}
\end{center}
\end{figure}

\begin{figure}[h]
\begin{center}
\includegraphics[height=.2\textheight]{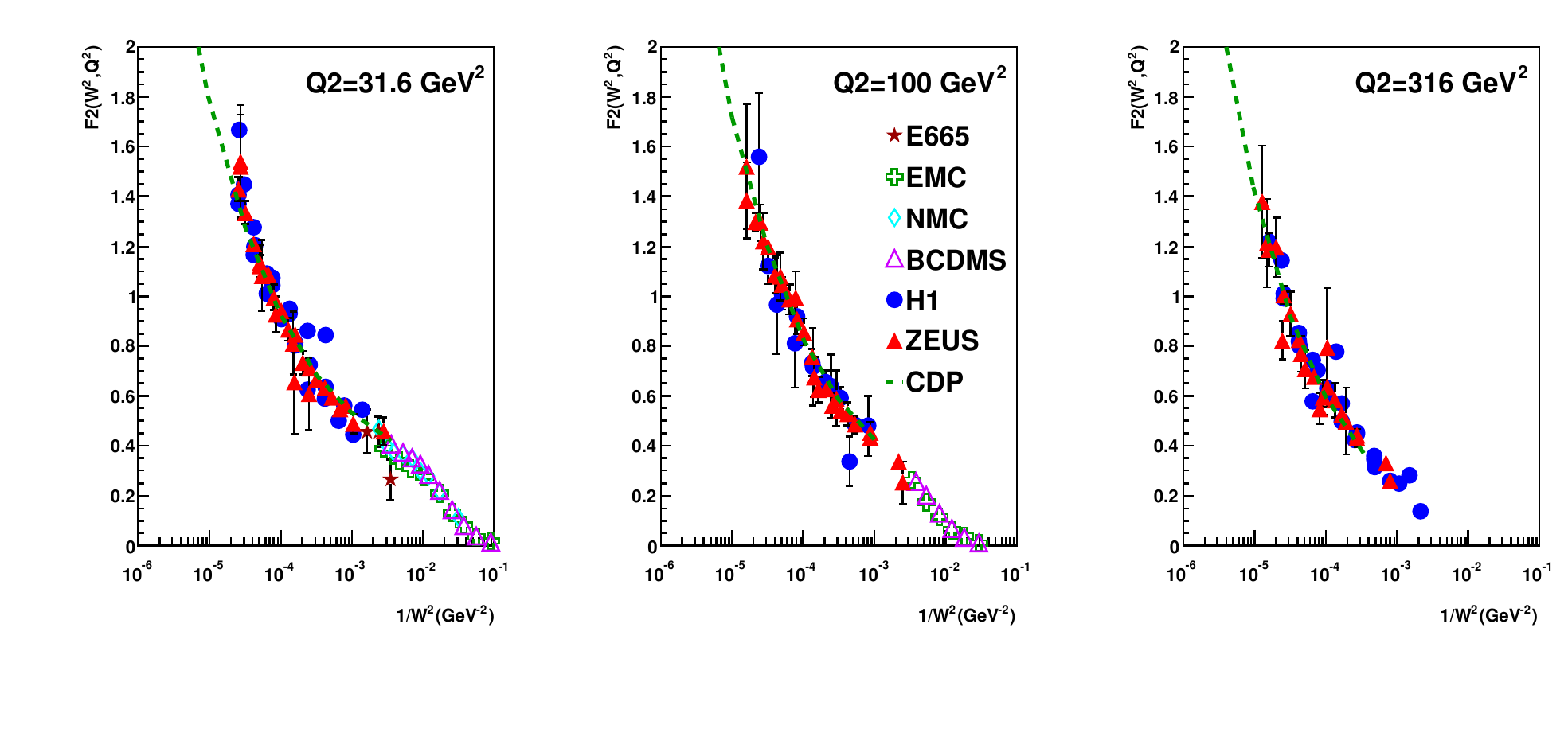}
\vspace*{-0.5cm}
\caption{{\footnotesize As in Fig.\,7, but for $31.6~{\rm GeV}^2 
\le Q^2 \le 316~{\rm GeV}^2$.}}
\end{center}
\end{figure}

\vspace*{-0.8cm}

\begin{center}{\bf CONCLUSIONS}\end{center}

\noindent
Scaling in $\eta(W^2,Q^2)$ of the total photoabsorption cross section, as 
well as the specific functional
dependences on $\eta (W^2,Q^2)$, corresponding to either
color transparency or saturation, have been recognized as general consequences
from
the color-gauge-invariant interaction of  dipole
fluctuations of the photon, $\gamma^* \to q \bar q$,
with the gluon field in the nucleon. Color
transparency corresponds to cancellation between the two interaction channels
related to the two diagrams in Fig. 5. The vanishing of this cancellation
in the high-energy limit of $W^2 \to \infty$ at any fixed value of $Q^2$
implies saturation of the cross section to the photoproduction limit.

\begin{center}{\bf ACKNOWLEDGMENTS}\end{center}

\noindent
Many thanks to Roberto Fiore, Alessandro Papa and
Agustin Sabio Vera for the
invitation to Diffraction 2012 and the efficient organization of a very
successful workshop in the beautiful surroundings of Lanzarote.

\end{document}